\documentclass[nofootinbib,preprint,aps,superscriptaddress,tight enlines,amsmath,amssymb]{revtex4}

\usepackage{bm}
\usepackage{epsfig}

\def\a{{\alpha}}
\def\b{{\beta}}
\def\d{{\delta}}
\def\D{{\Delta}}
\def\e{{\epsilon}}

\def\l{{\lambda}}
\def\L{{\Lambda}}

\def\S{{\Sigma}}
\def\s{{\sigma}}

\def\CPT{{$\chi$PT\ }}

\def\PQCPT{{PQ$\chi$PT\ }}

\def\str{\text{str}}

\def\cL{{\mathcal L}}

\def\cM{{\mathcal{M}}}
\def\fH{{\mathcal{H}}}

\def\OMIT#1{}

\newcommand{\nl}{\nonumber \\}
\newcommand{\nn}{\nonumber}

\newcommand{\bea}{\begin{eqnarray}}
\newcommand{\eea}{\end{eqnarray}}
\newcommand{\bsube}{\begin{subequations}}
\newcommand{\esube}{\end{subequations}}

\newcommand{\PQ}{\mbox{\tiny PQ}}

\newcommand{\bfb}[1]{\mbox{\boldmath $ #1 $}}

\newcommand{\Eq}[1]{Eq.\,(\ref{#1})}

\begin{document}

 \title{Chiral corrections to heavy quark-diquark
 symmetry predictions
 for doubly heavy baryon zero-recoil semileptonic decay}
  \author{Jie Hu} \email{hujie@ust.hk}
   \affiliation{Department of Physics, Duke University, Durham, North Carolina 27708, USA}
   \affiliation{Department of Chemistry,
  Hong Kong University of Science and Technology,
  Kowloon, Hong Kong}
  \date{\today}

\begin{abstract}
This paper studies the leading chiral corrections to heavy
quark-diquark symmetry predictions for doubly heavy baryon
semileptonic decay form factors. We derive the coupling between heavy
diquarks and weak current in the limit of heavy quark-diquark
symmetry, and construct the chiral Lagrangian for doubly heavy
baryons coupled to weak current. We evaluate chiral corrections to
doubly heavy baryon zero-recoil semileptonic decay for both
unquenched and partially quenched QCD. This theory is used to
derive chiral extrapolation expressions for measurements of form
factors of doubly heavy baryon zero-recoil semileptonic decay in
lattice QCD simulations.
\end{abstract}


  \maketitle

\section{Introduction}

This work is stimulated by the recent SELEX experimental
observation on states which have been tentatively interpreted as
doubly charm
baryons~\cite{Mattson:2002vu,Moinester:2002uw,Ocherashvili:2004hi}.
The masses and hyperfine splittings of the observed states
are consistent with theoretical predictions from some quark
models~\cite{Ebert:2002ig} and quenched lattice quantum
chromodynamics (QCD)
~\cite{Lewis:2001iz,Mathur:2002ce,Flynn:2003vz}.
However many other aspects are difficult to understand
in the current theoretical framework, such as the missing weak decay signal
of the higher state in the ground doublet.
 Model-independent predictions for doubly
heavy baryon properties would be helpful.
Chiral and heavy quark
symmetries are useful approximate symmetries of QCD for predicting
low-energy properties of hadrons containing a single heavy quark.
Heavy quark-diquark symmetry relates the properties of heavy mesons
with a single heavy quark ($Q\bar q$) to those of doubly heavy
anti-baryons with two heavy anti-quarks ($\bar Q \bar Q \bar q$).
The heavy quark-diquark symmetry prediction for the
doubly charm baryon hyperfine
splitting~\cite{Brambilla:2005yk,Fleming:2005pd,Lewis:2001iz,Ebert:2002ig}
is within $25 \sim 30\%$ of the quenched lattice QCD
calculation. More observables are needed to see whether the heavy
quark-diquark symmetry is applicable to charm and bottom.
Semileptonic decay of doubly charm baryons was
studied previously using model dependent method~\cite{Ebert:2004ck}
and recently by the heavy quark spin symmetry~\cite{Flynn:2007qt}.
However the latter provided only tree level predictions of
doubly heavy baryon semileptonic decay matrices.

In this paper we will study the chiral corrections to heavy
quark-diquark symmetry predictions for the form factors of zero-recoil
doubly heavy baryon semileptonic decay, and further arrive at
 the results for both unquenched and partially quenched QCD.

We use the techniques in non-relativistic quantum
chromodynamics (NRQCD) to derive the coupling
between heavy diquarks and weak current.
Savage and Wise~\cite{Savage:1990di} wrote an
effective Lagrangian for heavy quarks and diquarks using heavy
quark effective theory (HQET)~\cite{Manohar:2000dt} and calculated
the doubly heavy baryon hyperfine splittings. HQET is formulated
as an expansion in the heavy quark mass scaled $\Lambda_{QCD} /
m_Q$. Hadrons with two or more heavy quarks, such as quarkonia or
doubly heavy baryons, are characterized by the additional scales
$m_Q v$ and $m_Q v^2$, where $v$ is the typical velocity of the
heavy quark within the bound state. The dynamics of hadrons with
two or more heavy quarks follows
NRQCD~\cite{Bodwin:1994jh,Brambilla:1999xf,Luke:1999kz}. Within
this framework the effective Lagrangian for heavy diquark have
been derived recently~\cite{Brambilla:2005yk, Fleming:2005pd},
which corrected the hyperfine splitting predictions of
Ref.~\onlinecite{Savage:1990di} by a factor of $2$.

For studying the low-energy dynamics of hadrons containing
diquark, it is useful to build an effective theory which
incorporates the relevant symmetries of QCD. Heavy hadron chiral
perturbation theory
(HH$\chi$PT)~\cite{Wise:1992hn,Burdman:1992gh,Yan:1992gz} has
heavy hadrons, Goldstone bosons, and photons as its elementary
degrees of freedom and incorporates chiral and heavy quark
symmetries. Hu and Mehen~\cite{Hu:2005gf} derived an effective
Lagrangian for doubly heavy baryons incorporating heavy
quark-diquark symmetry and used this theory to study strong and
electromagnetic interactions of doubly heavy baryons in low-energy
regime. Here we will apply HH$\chi$PT to evaluate  chiral
corrections to semileptonic decay at zero-recoil.

Moreover, we will extend the
chiral Lagrangian with heavy quark-diquark symmetry to include lattice
effects of partial quenching.
We derive
expressions for the chiral extrapolation of zero-recoil
semileptonic decay of double heavy baryons in lattice calculation.
Lattice calculations of decay matrix
elements of heavy mesons and doubly heavy baryons can help
determine the reliability of heavy quark-diquark symmetry for
charm and bottom hadrons. High-precision lattice calculation
of semileptonic decay form factors of heavy mesons is a long term
project of several collaborations, such as HPQCD and MILC. Lattice
QCD simulations start with unphysical sea quark masses which need
to be extrapolated to physical values. Chiral Lagrangians in
quenched~\cite{Sharpe:1992ft} and partially
quenched~\cite{Sharpe:1997by, Sharpe:2001fh}
theory are useful for the chiral
extrapolations of the lattice simulation data.

The rest of the paper is organized as follows. In
Sec.~\ref{nrqcd}, we use NRQCD to derive the coupling between heavy
diquarks and weak current consistent with heavy quark-diquark
symmetry. In Sec.~\ref{lag}, we construct the chiral Lagrangian
for doubly heavy baryons coupled to weak current and calculate the
doubly heavy baryon zero-recoil semileptonic decay matrix
elements. We evaluate  the chiral corrections to the decay form factors
 using $\chi$PT. In the heavy
quark-diquark limit the corrections vanish. In Sec.~\ref{pq}, we
extend the chiral Lagrangian with heavy quark-diquark symmetry to
 partially quenched theory and derive form factor formulism for
the lattice QCD chiral extrapolation of doubly heavy baryon
zero-recoil semileptonic decay, via partial quenching versus unquenching.
Summary is given in Sec.~\ref{sum}. Some useful formulations are
collected in Appendix.

\section{Coupling between Heavy Diquarks and Weak
Current from $v$NRQCD} \label{nrqcd}
NRQCD is the nonrelativistic effective theory for the dynamics of
heavy quarks. In NRQCD the important scales are among the four
quantities: the heavy quark mass  $m_Q$, the typical momentum $m_Q
v$ of heavy quarks within bound state, the typical kinetic energy
$m_Q v^2$ of heavy quarks, and $\Lambda_{QCD}$. The total momentum
of heavy quark field is taken to be the sum of the labeling
momentum $\mathbf p (\sim m_Q v)$ and the residual momentum
$\mathbf k (\sim m_Q v ^2)$. In $v$NRQCD~\cite{Fleming:2005pd}, which
has a consistent $v$ expansion, the Lagrangian for $ \bar Q \bar Q$
($\bar Q = \bar b , \bar c$) of the leading order reads
\begin{align} \label{vnrqcd}
 \mathcal{L}
&= -\frac{1}{4}F^{\mu\nu}F_{\mu\nu}
\nl&\quad
 + \sum_{f=b,c} \sum_{\mathbf{p}} \chi_{\mathbf{p}}^ {f\dag}
  \Big( i D_0 - \frac{(\mathbf{p}- i \mathbf{D})^2}{2 m_{Q_f}}
    + \frac{g_s}{2 m_{Q_f}}
  {\bm\sigma} \cdot \mathbf{B}\Big)
     \chi_{\mathbf{p}}^ {f}
\nl&\quad
 - \frac{1}{2}\sum_{f=b,c} \sum_{\mathbf{p,q}}
   \frac{g_s^2}{(\mathbf{p}- \mathbf{q})^2}
    \chi_{\mathbf{q}}^{f\dag}\bar{T}^A
    \chi_{\mathbf{p}}^{f}\chi_{-\mathbf{q}}^{f\dag}\bar{T}^A
    \chi_{-\mathbf{p}}^ {f}
\nl&\quad
 - \sum_{\mathbf{p,q}} \frac{g_s^2}{(\mathbf{p}-\mathbf{q})^2}
   \chi_{\mathbf{q}}^{b\dag}\bar{T}^A
   \chi_{\mathbf{p}}^ {b}\chi_{-\mathbf{q}}^{c\dag}\bar{T}^A
   \chi_{-\mathbf{p}}^ {c} + \cdots .
\end{align}
Here $\chi_{\mathbf p}^f$ is a nonrelativistic anti-quark field
which annihilates an anti-quark of flavor $f$, $\bar T^A$ is
$SU(3)$ color generator for $\bar 3$ representation, $\mathbf{B}$
is chromomagnetic field, and $D_0$ and $\mathbf{D}$ are the time
and spatial components of the gauge covariant derivative,
respectively. The kinetic energy $D_0$ and momentum $\mathbf p$ of
heavy anti-quark field are of $O(m_Q v^2)$ and $O(m_Q v)$, respectively.

In order to derive an effective Lagrangian for diquark we follow the
methods of Ref.~\onlinecite{Fleming:2005pd}. We use the spin and
color Fierz identities
\bsube \label{spincolor},
\begin{align}
 \delta_{\alpha \delta} \delta_{\beta \gamma}
&= - \frac{1}{2}(\sigma ^i \epsilon)_{\alpha \beta}
    (\epsilon \sigma ^i)_{\gamma \delta}
    +\frac{1}{2} \epsilon_{\alpha \beta} \epsilon ^T_{\gamma \delta}\,,
\label{spin} \\
 \bar{T}^A_{il}\bar{T}^A_{jk}
&= \frac{2}{3} \sum_m \frac{1}{2} \epsilon_{mij} \epsilon_{mkl}
 + \frac{1}{3} \sum_{(mn)}d^{(mn)}_{ij}d^{(mn)}_{kl} ,
\label{color}
\end{align}
\esube
respectively, to decompose the
anti-quark bilinear such as $\chi_{\mathbf{p}}^ {f}
\chi_{\mathbf{-p}}^ {f}$ into operators with spin $0$ and spin
$1$, and to project the potential onto the
$\mathbf{3}$ and $\bar{\mathbf{6}}$ channels.
In \Eq{spincolor} the Greek subindexes denote spins and
the Roman subindexes denote colors.
$\sigma^i$ are  Pauli matrices, $\epsilon = i
\sigma^2$, and  $d^{(mn)}_{ij}$ are elements of symmetric
matrices in color space defined by
\begin{equation}
d^{(mn)}_{ij} = \left\{
\begin{array}{cc}
(\delta_{mi} \delta_{nj} + \delta_{ni} \delta_{mj} ) /\sqrt{2},
 & m\neq n \\
\delta_{mi} \delta_{nj}, & m = n
\end{array} \right. .
\end{equation}
The diquark states, $\bar b \bar b$ and $\bar c \bar c$,
must be only in
$(\mathbf{3})_C(\mathbf{3})_S $
or $(\bar{\mathbf{6}})_C(\mathbf{1})_S $
by the Pauli principle, while $\bar b \bar c$ can
also be in $(\mathbf{3})_C(\mathbf{1})_S $
or $(\bar{\mathbf{6}})_C(\mathbf{3})_S $.
By fourier transforming
$\frac{g_s^2}{(\mathbf{p}- \mathbf{q})^2}=
 \int d^3 \mathbf{r}\frac{g_s^2}{4 \pi r}
 e^{i (\mathbf{p}-\mathbf{q}) \cdot \mathbf{r}}$,
 \Eq{vnrqcd} leads to
\begin{widetext}
\begin{align}\label{lagran}
\mathcal{L} &=-\frac{1}{4}F^{\mu\nu}F_{\mu\nu}
+ \sum_{f=b,c} \sum_{\mathbf{p}}
\chi_{\mathbf{p}}^ {f\dag}
\Big( i D_0 - \frac{(\mathbf{p}- i \mathbf{D})^2}{2 m_{Q_f}}
+ \frac{g_s}{2 m_{Q_f}} {\bm \s} \cdot \mathbf{B}\Big)
\chi_{\mathbf{p}}^ {f}
\nl & \quad
- \frac{1}{2} \sum_{f=b,c} \int d^3 \mathbf{r} V^{(3)}(r)
\Big( \sum_{\mathbf{q}}
e^{-i \mathbf{q} \cdot \mathbf{r}} \frac{\epsilon_{mij}}{2}
(\chi_{\mathbf{q}}^ {f\dag})_i {\bm \s} \epsilon
(\chi_{\mathbf{-q}}^ {f\dag})_j \Big)
\cdot \Big( \sum_{\mathbf{p}} e^{i \mathbf{p} \cdot \mathbf{r}}
\frac{\epsilon_{mkl}}{2}
(\chi_{\mathbf{-p}}^ {f})_k \epsilon {\bm \s}
 (\chi_{\mathbf{p}}^ {f})_l \Big)
\nl &\quad
- \int d^3 \mathbf{r} V^{(3)}(r)
 \Big( \sum_{\mathbf{q}} e^{-i \mathbf{q} \cdot \mathbf{r}}
\frac{\epsilon_{mij}}{2} (\chi_{\mathbf{q}}^ {b\dag})_i {\bm \s}
\epsilon
(\chi_{\mathbf{-q}}^ {c\dag})_j \Big) \cdot \Big( \sum_{\mathbf{p}}
e^{i \mathbf{p} \cdot \mathbf{r}}
 \frac{\epsilon_{mkl}}{2} (\chi_{\mathbf{-p}}^ {c})_k
\epsilon {\bm \s}  (\chi_{\mathbf{p}}^ {b})_l \Big)
\nl &\quad
- \int d^3 \mathbf{r} V^{(3)}(r)
\Big(- \sum_{\mathbf{q}} e^{-i \mathbf{q} \cdot \mathbf{r}}
 \frac{\epsilon_{mij}}{2} (\chi_{\mathbf{q}}^ {b\dag})_i  \epsilon
 (\chi_{\mathbf{-q}}^ {c\dag})_j \Big)
\Big( \sum_{\mathbf{p}} e^{i \mathbf{p} \cdot \mathbf{r}}
\frac{\epsilon_{mkl}}{2} (\chi_{\mathbf{-p}}^ {c})_k
\epsilon^T (\chi_{\mathbf{p}}^ {b})_l \Big)
 + \int d^3 \mathbf{r} V^{(\bar 6)}(r) [\cdots] ,
\end{align}
\end{widetext}
where $ V^{(3)}(r) = -\frac{2}{3}\frac{\alpha_s}{r}$.
The last term with  $V^{(\bar 6)}(r) =
\frac{1}{3}\frac{\alpha_s}{r}$ is irrelevant to
color neutral doubly heavy baryon and is therefore
dropped out hereafter.
The first term is the gauge boson field.
The second term includes the kinetic operators for the
anti-quark fields and the leading
spin symmetry breaking interactions which generate hyperfine
splittings. The other terms are the quartic terms with
diquark fields.

The diquark fields for $\bar{c} \bar{c}$ and
$\bar{b} \bar{b}$ can be introduced using the
Hubbard-Strantonovich transformation as
Ref.~\onlinecite{Fleming:2005pd}.
In addition, we introduce diquark field of
different flavor $\bar b \bar c$
to get the interaction for doubly heavy
baryon semileptonic decay.
We add to the Lagrangian in \Eq{lagran} the
following identity of $\Delta \mathcal{L}=0 $:
\begin{widetext}
\begin{align} \label{deltal}
\Delta \mathcal{L}&= \frac{1}{2}\sum_{f=b,c}
\int d^3 \mathbf{r} V^{(3)}(r)\Big(\!\mathbf{T}^{m
\dag}_{\mathbf r}\! -\!\sum_{\mathbf{q}}\! e^{-i \mathbf{q} \cdot
\mathbf{r}} \frac{\epsilon_{mij}}{2} (\chi_{\mathbf{q}}^
{f\dag})_i {\bm \s} \epsilon
(\chi_{\mathbf{-q}}^ {f\dag})_j\!\Big)
 \!\cdot\! \Big(\!\mathbf{T}^{m}_{\mathbf r}\! -\!\sum_{\mathbf{p}}\!
e^{i \mathbf{p} \cdot \mathbf{r}}
 \frac{\epsilon_{mkl}}{2} (\chi_{\mathbf{-p}}^ {f})_k
\epsilon {\bm \s} (\chi_{\mathbf{p}}^ {f})_l\!\Big)\!
\nl &\quad
+\!\int\! d^3 \mathbf{r} V^{(3)}(r)\!
\Big(\!\tilde{\mathbf{T}}^{m \dag}_{\mathbf r}\!
-\!\sum_{\mathbf{q}}\! e^{-i \mathbf{q} \cdot \mathbf{r}}
\frac{\epsilon_{mij}}{2}
(\chi_{\mathbf{q}}^ {b\dag})_i {\bm \s} \epsilon
(\chi_{\mathbf{-q}}^ {c\dag})_j\!\Big)
\!\cdot\! \Big(\!\tilde{\mathbf{T}}^m_{\mathbf r}\!
-\!\sum_{\mathbf{p}}\!
e^{i \mathbf{p} \cdot \mathbf{r}} \frac{\epsilon_{mkl}}{2}
 (\chi_{\mathbf{-p}}^ {c})_k
\epsilon {\bm \s} (\chi_{\mathbf{p}}^ {b})_l\!\Big)\!
\nl &\quad
+\!\int\!d^3 \mathbf{r} V^{(3)}(r)\!\Big(\!T'^{m \dag}_{\mathbf r}
+\!\sum_{\mathbf{q}} e^{-i \mathbf{q} \cdot \mathbf{r}}
 \frac{\epsilon_{mij}}{2}
(\chi_{\mathbf{q}}^ {b\dag})_i \epsilon
 (\chi_{\mathbf{-q}}^ {c\dag})_j\!\Big)\!
\Big(\!T'^{m}_{\mathbf r}- \!\sum_{\mathbf{p}}
 e^{i \mathbf{p}\cdot \mathbf{r}}
 \frac{\epsilon_{mkl}}{2}
  (\chi_{\mathbf{-p}}^ {c})_k \epsilon^T
  (\chi_{\mathbf{p}}^ {b})_l\!\Big)\!,
\end{align}
where $m=1,2,3$ is the color index.
$\mathbf{T}^{m}= \sum_{\mathbf{p}}
e^{i \mathbf{p} \cdot \mathbf{r}}
\frac{\epsilon_{mkl}}{2} (\chi_{\mathbf{-p}}^ {f})_k
\epsilon {\bm \s} (\chi_{\mathbf{p}}^ {f})_l$,
$\tilde{\mathbf{T}}^m= \sum_{\mathbf{p}}
e^{i \mathbf{p} \cdot \mathbf{r}}
\frac{\epsilon_{mkl}}{2} (\chi_{\mathbf{-p}}^ {c})_k
\epsilon {\bm \s}
(\chi_{\mathbf{p}}^ {b})_l$,
and
$T'^m=\sum_{\mathbf{p}}
e^{i \mathbf{p} \cdot \mathbf{r}} \frac{\epsilon_{mkl}}{2}
(\chi_{\mathbf{-p}}^ {c})_k \epsilon^T
 (\chi_{\mathbf{p}}^ {b})_l$
annihilate
a spin-$1$ $\bar f \bar f$ field,
a spin-$1$  $\bar b \bar c$ field, and
a spin-$0$ $\bar b \bar c$ field, respectively.
The quartic terms of anti-quark fields in the
combination of \Eq{lagran} and \Eq{deltal}
are cancelled, resulting in
\begin{align} \label{interaction}
 \mathcal{L}
&= \frac{1}{2} \sum_{f=b,c} \int d^3 \mathbf{r}\,
  V^{(3)}(r) \Big(\mathbf{T}^{m \dag}_{\mathbf r}
   \mathbf{T}^{m}_{\mathbf r}-\mathbf{T}^{m \dag}_{\mathbf r}
   \sum_{\mathbf{p}} e^{i \mathbf{p} \cdot \mathbf{r}}
    \frac{\epsilon_{mkl}}{2} (\chi_{\mathbf{-p}}^ {f})_k \epsilon
    \bfb{\sigma} (\chi_{\mathbf{p}}^ {f})_l
  -\mathbf{T}^{m}_{\mathbf r} \sum_{\mathbf{q}}
  e^{-i \mathbf{q} \cdot \mathbf{r}}
   \frac{\epsilon_{mij}}{2} (\chi_{\mathbf{q}}^{f\dag})_i
   \bfb{\sigma}\epsilon (\chi_{\mathbf{-q}}^ {f\dag})_j
  \Big)
\nl&\quad
 +\int d^3 \mathbf{r} V^{(3)}(r)
   \Big(\tilde{\mathbf{T}}^{m\dag}_{\mathbf r}
   \tilde{\mathbf{T}}^{m}_{\mathbf r}
 -\tilde{\mathbf{T}}^{m\dag}_{\mathbf r}
   \sum_{\mathbf{p}} e^{i\mathbf{p} \cdot \mathbf{r}}
   \frac{\epsilon_{mkl}}{2}
   (\chi_{\mathbf{-p}}^{c})_k \epsilon
   \bfb{\sigma} (\chi_{\mathbf{p}}^ {b})_l
 -\tilde{\mathbf{T}}^{m}_{\mathbf r}
   \sum_{\mathbf{q}} e^{-i \mathbf{q} \cdot \mathbf{r}}
   \frac{\epsilon_{mij}}{2} (\chi_{\mathbf{q}}^{b\dag})_i
  \bfb{\sigma}\epsilon (\chi_{\mathbf{-q}}^{c\dag})_j\!
  \Big)
\nl &\quad
 + \int d^3 \mathbf{r} V^{(3)}(r)
  \Big(T'^{m \dag}_{\mathbf r} T'^{m}_{\mathbf r}
 - T'^{m \dag}_{\mathbf r}
     \sum_{\mathbf{p}} e^{i\mathbf{p} \cdot \mathbf{r}}
      \frac{\epsilon_{mkl}}{2}
      (\chi_{\mathbf{-p}}^{c})_k \epsilon^T (\chi_{\mathbf{p}}^ {b})_l
 + T'^{m}_{\mathbf r}\sum_{\mathbf{q}} e^{-i \mathbf{q} \cdot \mathbf{r}}
   \frac{\epsilon_{mij}}{2}
  (\chi_{\mathbf{q}}^{b\dag})_i \epsilon (\chi_{\mathbf{-q}}^ {c\dag})_j
 \Big).
\end{align}
\end{widetext}

\begin{figure}
\includegraphics*[width=0.9\columnwidth]{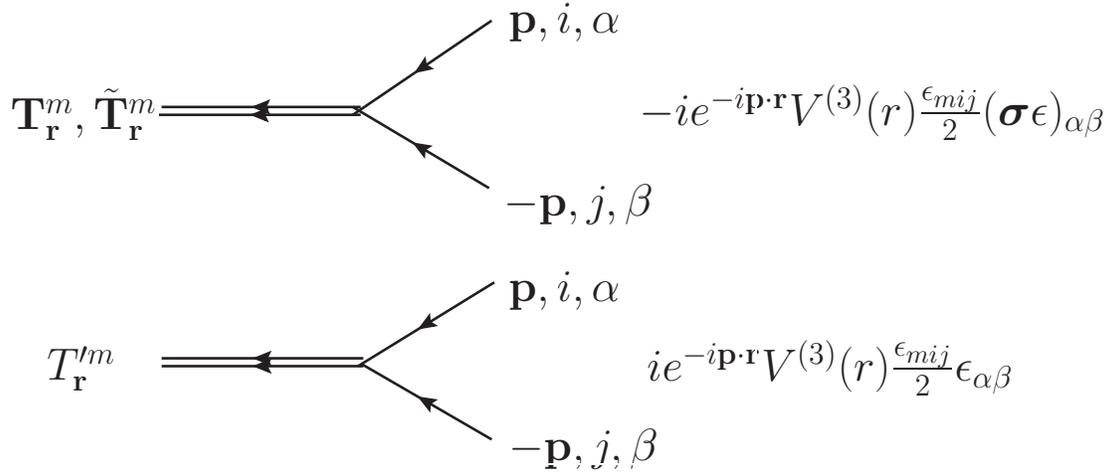}
\caption{Feynman rules for the coupling between the diquark
($\mathbf{T}_{\mathbf r}^m, \tilde{\mathbf T}_{\mathbf r}^m $ or
$T'^m_{\mathbf r}$) and anti-quarks. The Greek indexes denote spins
and the Roman indexes refer to colors.}
\label{feynman}
\end{figure}

Integrating out the fields
$\mathbf T^m_{\mathbf r}$, $\tilde{\mathbf T}^m_{\mathbf r}$ and
$T'^m_{\mathbf r}$ in \Eq{interaction} recovers the original NRQCD Lagrangian;
integrating out the anti-quark fields $\chi ^b_{\mathbf p}$ and
$\chi ^c_{- \mathbf p}$ yields an effective action for the diquark
fields $\mathbf T^m_{\mathbf r}$, $\tilde{\mathbf T}^m_{\mathbf r}$ and
$T'^m_{\mathbf r}$.
The kinetic terms of $\bar b \bar c$ fields are the
same as those of $\bar c \bar c$ fields in Ref.~\onlinecite{Fleming:2005pd},
with the proper diquark reduced mass
$ \mu_Q = (m_b +m_c)/(m_b m_c)$.
The Feynman rules for the coupling between the diquark field and the pair
of anti-quarks in \Eq{interaction} are
shown in Fig.~\ref{feynman}.

\begin{figure}
\centering
\epsfig{file=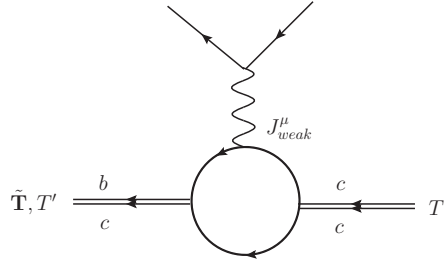, width=65mm}
 {\tighten
\caption[1]{One loop diagram contributing to the coupling between
composite anti-diquark fields and weak current.} }
\label{loop}
\end{figure}

Flavor changing weak current is given by
$J^{\mu}_{QCD} = \bar c \gamma ^{\mu} (1- \gamma_5)b$
and is rewritten as $J^{\mu}_{NR} =
\chi^{c\dag}(\delta_{\mu 0}- \delta_{\mu i} \sigma^i ) \chi^b$
by mapping onto NRQCD to the lowest order.
The coupling between anti-diquarks and weak current is to be evaluated
by the lowest order diagram as in Fig.~\ref{loop}
and the resulting Lagrangian for weak interactions is
\begin{align}  \label{diweak}
\mathcal{L}_{\rm weak} &= J^\mu_{\bar Q \bar Q} (J_{\rm weak})_\mu ,
\nl &
=- \int d^3 \mathbf{r} \big[\delta_{\mu 0} \mathbf{T} ^{i\dag}_{\mathbf r}
\cdot \tilde{\mathbf{T}} ^{i}_{\mathbf r} + i \delta_{\mu n}
(\mathbf{T}^{i\dag}_{\mathbf r} \times \tilde{\mathbf{T}} ^{i}_{\mathbf r} )^n
\nl & \quad
+ \delta_{\mu n} (T ^{i\dag}_{\mathbf r})^n  T'^{i}_{\mathbf r} \big]
(J_{\rm weak})_\mu ,
\nl &
\approx -  \eta \big[\delta_{\mu 0} \mathbf{T} ^{i\dag} \cdot
\tilde{\mathbf{T}} ^{i} + i \delta_{\mu n}
(\mathbf{T} ^{i\dag} \times \tilde{\mathbf{T}} ^{i} )^n
\nl & \quad
+\delta_{\mu n} (T ^{i\dag})^n  T'^{i} \big] (J_{\rm weak})_\mu,
\end{align}
where $(J_{\rm weak})_\mu$ is the weak current and $J^\mu_{\bar Q \bar Q}$ is the
diquark current. In the last identity we expanded the diquark fields
to the lowest order and wrote the Lagrangian in terms of local
current.  The factor $\eta$ can be interpreted as the
spatial wavefunction overlap of the ground state initial diquark
with the ground state final diquark system, which is not predicted
by symmetry, i.e. $\eta =  \int d^3 \mathbf{r} \phi^{\ast}_{\bar c\bar c}(\mathbf
r) \phi_{\bar b\bar c}(\mathbf r)$.

\section{Doubly Heavy Baryon Semileptonic Decay} \label{lag}
In the heavy quark limit, the ground states of the spin-$0$ heavy meson
$P$ and the spin-$1$ heavy meson $\mathbf{P}^{\ast}$ are degenerate, and therefore
combined into a single field,
\bea
H_{a, \a\b} = (\mathbf{P}^{\ast}_a \cdot {\bm \sigma})_{\a\b} + P_{a,\a\b}  \, ,
\eea
where $a$ is  $SU(3)$ flavor anti-fundamental index and ${\bm \sigma}$
is the vector of Pauli matrices.
Similarly in the heavy quark limit,
the doubly heavy anti-baryon ground state doublet $T_{i \beta}$
consists of a
spin-$\frac{1}{2}$ doubly heavy anti-baryon $\Xi_{a,\gamma}$ and a
spin-$\frac{3}{2}$ doubly heavy anti-baryon $\Xi^{\ast}_{a, i \beta}$ as
\bea\label{tsf}
T_{a,i \beta} = \sqrt{2}\Big(\Xi^{\ast}_{a,i \beta}+ \frac{1}{\sqrt{3}}
\Xi_{a,\gamma} \,\sigma^i_{\gamma \beta} \Big) \, .
\eea
Hereafter $T$ denotes the doubly heavy anti-baryon field
rather than the anti-diquark field in the previous section.
In \Eq{tsf} the index of diquark spin runs as $i=1,2,3$ and
that of light anti-quark spin as $\beta = 1,2$. The field
$\Xi^{\ast}_{a,i \beta}$ obeys the constraint
$\Xi^{\ast}_{a,i \beta} \, \sigma^i_{\beta\gamma}=0$.
Heavy quark-diquark symmetry relates properties of
heavy mesons to those of doubly heavy anti-baryons.
By the symmetry the effective Lagrangian for doubly heavy anti-baryon
and heavy meson ground
state doublets in the heavy hadron rest frame was
constructed in Ref.~\onlinecite{Hu:2005gf},
\begin{align}\label{qdql}
\mathcal {L} &= {\rm Tr}[{\cal H}^\dagger_a (i D_0)_{ba} {\cal H}_b]
- g \, {\rm Tr}[{\cal H}^\dagger_a  {\cal H}_b \, {\bm \sigma}\cdot {\mathbf A}_{ba}]
\nl & \quad
+\frac{\Delta_H}{4}{\rm Tr}[{\cal H}^\dagger_a \,  \Sigma^i \, {\cal H}_a \, \sigma^i],
\end{align}
where ${\Sigma}^i = \{{\Sigma}^i_{\mu \nu};~\mu,\nu=1,\cdots,5\}$ with matrix
elements
${\Sigma}^i_{\mu \nu}= {\sigma}^i_{\alpha \beta}\delta_{\mu \a}\delta_{\nu\b}
 -i \epsilon_{ijk}\delta_{\mu,j+2}\delta_{\nu,k+2}$.
Here ${\cal H}_{a}=\{{\cal H}_{a,\mu \beta};~\mu= 1, \cdots,5 $ and $\b =1,2\}$
is a $5 \times 2$
matrix field, with elements
${\cal H}_{a,\mu \beta} = H_{a,\alpha \beta}\delta_{\mu \alpha}
+ T_{a,i \beta}\delta_{\mu,i+2}$.
It transforms as tensor product
of a five component field and a two-component
light anti-quark spinor. The five component field  corresponds to the two heavy
quark spin states and the three anti-diquark spin states.
In \Eq{qdql} the first term is the kinetic operator and
the second term is the coupling with the
axial current vector ${\mathbf A}_{ba}$.
These two terms respect the $SU(5)$ heavy quark-diquark
symmetry. The third term is the leading heavy quark symmetry breaking
operator and responsible to hyperfine splitting.
This term is characterized by charm meson hyperfine splitting $\Delta_ H$,
which is also $4/3$ of that of doubly charm
baryon by heavy quark-diquark symmetry.

For our interests we retain only the anti-baryon terms in \Eq{qdql}, i.e.,
\begin{align}\label{lagrangian}
 \mathcal{L}&= \tilde T^+ _ {\beta i} (i D_0) \tilde T_{i \beta}
 - g \tilde T^+ _ {\beta i} \tilde T_{i \beta '}
  {\sigma}^i _{\beta ' \beta}  A^i
 \nl & \quad
 + \Big(\frac{m_c / 2}{\mu_Q}\Big) \frac{\Delta _ H}{4}
 \tilde T^+ _ {\beta j} (-i \epsilon _{ijk}) \tilde T_{k \beta '}
 \sigma ^ i _{\beta ' \beta}
 \nl & \quad
+ T^+ _ {\beta i} (i D_0) T_{i \beta} - g T^+ _ {\beta i}
T_{i \beta '} \sigma ^i_{\beta ' \beta} A^i
 \nl & \quad
+  \frac{\Delta _ H}{4} T^+ _ {\beta j} (-i \epsilon _{ijk})
T_{k \beta '} \sigma ^ i _{\beta ' \beta}
 \nl & \quad
+ T'^+ _ {\beta } (i D_0) T'_{\beta} - g' T'^+ _ {\beta }
T'_{\beta '} \sigma^i _{\beta ' \beta} A^i \, .
\end{align}
Here, we have suppressed the flavor index $a$.
The composite fields $T$ are decomposed
in terms of spin-$\frac{3}{2}$ field $\Xi ^{\ast}_{i \beta} $
and spin-$\frac{1}{2}$  field $\Xi _\alpha$,
\bsube \label{tfield}
\begin{align}
T_{i \b ; \bar c \bar c \bar q} &= \sqrt{2}
\Big(\Xi_{ i \beta ;\bar c \bar c \bar q}^{\ast}
+ \frac{1}{\sqrt{3}} \Xi_{\alpha ; \bar c \bar c \bar q}
\sigma_{\alpha \beta} ^ i\Big), \\
 \tilde{T}_{i \beta ; \bar b \bar c \bar q}
 &= \sqrt{2} \Big(\Xi_{ i \beta ; \bar b \bar c \bar q}^{\ast}
 + \frac{1}{\sqrt{3}} \Xi_{ \alpha ; \bar b \bar c \bar q}
 \sigma_{\alpha \beta} ^ i\Big),\\
 T'_{\beta ; \bar b \bar c \bar q} &= \sqrt{2} \;
 \Xi_{\beta ; \bar b \bar c \bar q}' \, .
\end{align}
\esube
$T_{i \beta}$ and
$\tilde T_{i \beta} $ are the ground state doublet of $\bar c\bar
c \bar q$ and $\bar b\bar c \bar q$, respectively, with the diquark
spin-$1$. $T'_\beta$ is the spin-$\frac{1}{2}$ ground state
of $\bar b\bar c \bar q$ with diquark spin-$0$. The
hyperfine splittings of the doubly heavy anti-baryons are related to
those of the charm mesons by heavy quark-diquark symmetry,
$m_{\Xi^{\ast}_{cc}}-m_{\Xi_{cc}} = \frac{3}{4} \Delta_ H$ and
$m_{\Xi^{\ast}_{bc}}-m_{\Xi_{bc}} = \Big(\frac{m_c / 2}{\mu_Q}\Big)
\frac{3}{4} \Delta_ H$, where  $\mu_Q$ is the reduced mass of
diquark $bc$.

Equation(\ref{lagrangian}) describes strong and
electromagnetic interactions in the baryon rest
frame in the low-energy regime,
 in which the doubly heavy baryon four-velocity is conserved
(up to $O(\Lambda_{QCD}/m_Q$) corrections).
For a process such as  weak decay, in which the initial and
final baryons have different four-velocities, the covariant
representation of baryon field is needed. However, for
studying the zero-recoil semileptonic decay, in which the
doubly heavy baryon four-velocity is conserved, it is possible
to work in the baryon rest frame. Therefore it allows to
map the weak current coupling of the diquark in \Eq{diweak} onto a current
operator  for doubly heavy baryons as in Ref.~\onlinecite{Hu:2005gf}
and the resulting Lagrangian for semileptonic decays of doubly heavy baryon is
\begin{equation} \label{weaklag}
 \mathcal{L}_{\rm weak}= i \eta T^+ _ {\beta i} ( \epsilon _{ijk}
 \tilde T_{k \beta } J^k _{\rm weak}
  +i \tilde T_{i \beta} J^0 _{\rm weak}
  +  T'_{ \beta } J^i _{\rm weak} )\, .
\end{equation}
The coupling between doubly heavy baryons and weak current is
obtained by demanding that the anti-diquark index on the doubly heavy
baryons couple to the weak current as in Eq.(\ref{diweak}).

For Eq.(\ref{weaklag}) being written explicitly
in terms of doubly heavy anti-baryon
fields, we evaluate the weak current matrix elements  as
\begin{align}\label{formfactor}
\langle \Xi_{\a ' ; \bar c \bar c \bar q}|J^{\mu}_{\rm weak}|
\Xi_{\a ; \bar b \bar c \bar q}\rangle
&= \eta \bar{u}_{\alpha '}
\Big(\! -\!2i (1+\d^q_1)\, \delta _{\alpha \alpha '}
\delta _{\mu 0}
\nl & \quad
 - \tfrac{4}{3}i (1+\d^q_2)\,
\sigma ^j _{\alpha \alpha '} \delta _{\mu j} \Big)u_{\alpha } ,
\nl
\langle \Xi_{i \b ; \bar c \bar c \bar q}^{\ast}|J^{\mu}_{\rm weak}|
  \Xi_{\a ; \bar b \bar c \bar q}\rangle
&= \eta \bar{u}_{i \beta}
  \Big(  \tfrac{2}{\sqrt{3}}i (1+\d^q_3)\,\delta_{\alpha \beta}
  \delta _{\mu i}\Big)u_{\alpha } ,
\nl
\langle\Xi_{\a ; \bar c \bar c \bar q}|J^{\mu}_{\rm weak}|
  \Xi_{i \b ; \bar b \bar c \bar q}^{\ast}\rangle
&= \eta  \,
  \bar{u}_{\alpha}\Big( \tfrac{2 }{\sqrt{3}}i (1+\d^q_4)\,
  \delta_{\alpha \beta} \delta _{\mu i} \Big)u_{i \beta }  ,
\nl
\langle\Xi_{k \b ; \bar c \bar c \bar q}^{\ast}|J^{\mu}_{\rm weak}|
  \Xi_{i \a ; \bar b \bar c \bar q}^{\ast}\rangle
&=\eta
  \bar{u}_{ k \beta}\Big(\!-\!2i (1+\d^q_5)\,
  \delta _{\alpha \beta}\delta_{ik} \delta _{\mu 0}
\nl &\quad
  + 2 i (1+\d^q_6)\,\sigma ^j _{\alpha \beta}
  \delta_{ik} \delta _{\mu j}\Big)u_{i \alpha }  ,
\nl
\langle \Xi_{\a ' ; \bar c \bar c \bar q}|J^{\mu}_{\rm weak}|
  \Xi'_{\a ; \bar b \bar c \bar q}\rangle
&= \eta   \bar{u}_{\alpha '}
  \Big( \tfrac{-2}{\sqrt{3}} (1+\d^{q\prime}_1) \,
  \sigma ^j _{\alpha \alpha '} \delta _{\mu j} \Big)u'_{\alpha },
\nl
\langle \Xi_{i \b ; \bar c \bar c \bar q}^{\ast}|J^{\mu}_{\rm weak}|
  \Xi'_{\a ; \bar b \bar c \bar q}\rangle
&= \eta
  \bar{u}_{ i \beta}\Big( \!-\!2 (1+\d^{q\prime}_2)\,
  \delta_{\alpha \beta}  \delta _{\mu i}\Big)u'_{\alpha } ,
\end{align}
where $(u_\alpha, u'_\alpha, u_{i \beta})$ and
$(\bar u_\alpha, \bar u_{i \beta}, \bar u'_\alpha)$
are nonrelativistic spinors for initial and final states,
respectively.
Up to an overall normalization, the form
factors in \Eq{formfactor} are dictated only
by heavy quark-diquark symmetry.
The $\d^q_i$-parameters are corrections to the form factors and receive
contributions from such as
heavy quark symmetry breaking effects.
The tree level semileptonic
decay matrix elements
 agree with Ref.~\onlinecite{Flynn:2007qt}
up to an overall minus sign.
It is noted that they gave results for
$\Xi_{bc}$ while we report  $\Xi_{\bar b \bar c}$.
To obtain $\Xi_{bc}$,
the decomposition of the ground state doublet
is constructed as
$T_{i \beta} = \sqrt{2}\Big(\Xi^{\ast}_{i \beta} - \frac{1}{\sqrt{3}}
\sigma^i_{\beta \gamma} \Xi_{\gamma} \Big)$
 rather than those in \Eq{tfield}.

\begin{figure}
\includegraphics*[width=0.5\columnwidth]{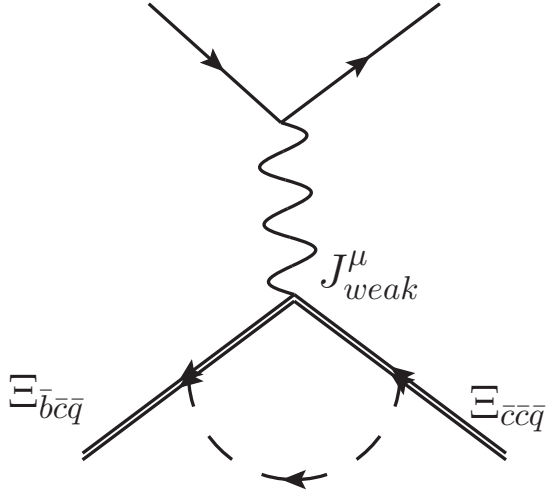}
\caption{One-loop contributions to the doubly heavy baryon semileptonic decay.}
\label{oneloop}
\end{figure}

At the tree level, the corrections to the form factors are zero and
in the heavy quark limit they vanish at any order.
The goal of this paper is to obtain those
heavy quark symmetry breaking corrections that come
from chiral loops such as in Fig.~\ref{oneloop}.
The  $\d^q_1$ to $\d^q_6$ parameters for
doubly heavy anti-baryons  are related to the known heavy
$D$ mesons through heavy quark-diquark symmetry. The
coupling constant $g$ in \Eq{lagrangian}
which appears in the chiral corrections
is available from $D$ meson measurements.
Chiral corrections as $\d^{q\prime}_1$ and $\d^{q\prime}_2$
depend on the coupling constant $g'$ that relates to
excited states and is yet
to be available experimentally.
We will only focus on corrections $\d^q_1$ to $\d^q_6$ for this paper.

The pion-baryon vertex in Fig.~\ref{oneloop} is generated
from the pion-baryon interaction term in the Lagrangian Eq.(\ref{lagrangian})
and the weak current vertex is from
Eq.(\ref{weaklag}). Loops for wavefunction
renormalization (not drawn here explicitly) are also contribute.
The chiral loop corrections to semileptonic decay form factors
are given in Appendix.
The partial quenching corrections will be discussed in the next section.

Table~\ref{tab:chicorr} provides numerical results for $\d^u_i$
and $\d^s_i$. Note that $\d^u_i = \d^d_i$ in the isospin limit.
\begin{table}
 \begin{center}
\begin{tabular}{|c|c|c|c|c|c|c|}
    \hline
    \hline
&$\d^u_1$ & $\d^u_2$& $\d^u_3$ &
$\d^u_4$& $\d^u_5$& $\d^u_6$\\
        \hline
$\mu =500$ MeV & 0.25 & 0.25 & 0.07 & 0.15 & -0.05 & -0.05\\
    \hline
$\mu =1500$ MeV & 0.32 & 0.32  & 0.10 & 0.19 & -0.01& -0.02\\
    \hline
    \hline
&$\d^s_1$ & $\d^s_2$& $\d^s_3$ & $\d^s_4$&
$\d^s_5$& $\d^s_6$\\
        \hline
$\mu =500$ MeV & -0.01 & -0.01 & -0.14  & -0.14  &-0.16&-0.16\\
    \hline
$\mu =1500$ MeV & -0.0002 & -0.001 & -0.13   & -0.12 &-0.13& -0.13 \\
    \hline
\end{tabular}
\caption{Chiral corrections to form factors of
doubly heavy baryon ($q=u, s$) semileptonic decay at $\mu = 500$
MeV and $\mu = 1500$ MeV. Other paraments see text.
Note that $\d^u_i = \d^d_i$.
}
\label{tab:chicorr}
\end{center}
\end{table}
For simplicity the reduced mass of diquark $bc$ is set to be
$\mu_Q = \frac{m_b m_c}{m_b+m_c} \approx m_c$.
In calculations, the parameters as chosen as follows.
$g = 0.6$~\cite{Fajfer:2006hi} and $\D_H = 140$ MeV are from
experiments;
Goldstone boson masses are set to be $m_\pi = 140$
MeV, $m_k = 500$ MeV,
and their decay constants are chosen to be equal $f_\pi = f_k =f_\eta = 130$
MeV.
$m_\eta = \sqrt{(- m_\pi ^2 +4 m_k ^2)/3}$ from $SU(3)$ prediction.
Then only one parameter, the renormalization scale $\mu$, can be varied.
In the chiral perturbation theory calculations, the logarithmic
$\mu$-dependence from loops is cancelled by $\mu$-dependent counterterms
which are not included here.
Therefore we vary $\mu$ from $500$ MeV to $1500$ MeV to obtain an estimate of
uncertainty due to the unknown counterterm contributions.
Table~\ref{tab:chicorr} shows results at $\mu=500$ MeV and $\mu= 1500$
MeV.
As shown the $\mu$ dependence is pretty small.

We find that the corrections to the decay form factors of
the doubly heavy baryon containing anti-up or
anti-down quark, $\d^u_1, \d^u_2$ in the
$\Xi_{bcu}^{1/2} \rightarrow \Xi_{ccu}^{1/2} $ transition get
chiral corrections of order $25 \sim 32\%$, while  $\d^u_5$,
$\d^u_6$ in the $\Xi_{bcu}^{3/2} \rightarrow \Xi_{ccq}^{3/2} $
transition get corrections of order $5\%$.
The form factors
for doubly heavy baryon containing anti-strange quark obtain
negative corrections and $\d^s_1,\d^s_2$ are least sensitive,
with corrections of order $1\%$. It would be
interesting to see if the observed deviations from heavy quark-diquark
symmetry in either experiments or lattice simulations agree with the
predictions from chiral perturbation theory as we shown.
If there is any disagreements with
$\chi$PT predictions, it will indicates
the $\L_{QCD}/m_Q$ heavy quark-diquark
corrections are dominated by short distance effects
which are not considered here.

\section{Partially quenched chiral corrections} \label{pq}

The effective field theory techniques
with heavy quark-diquark symmetry for
doubly heavy baryons can
also be applied to doubly heavy systems simulated on the lattice.
The $\chi PT$ results from the previous section are
extendable to include lattice
artifacts such as quenching and partial quenching.

In a lattice QCD simulation, the sea quark masses are often different
from the valence quark masses. There are mainly two ways of treating
sea quark masses. In partially quenched lattice QCD the sea quark
masses are different from the valence quark masses, while in
quenched lattice QCD, 
the sea quark contributions are absent.
To reproduce these lattice artifacts in effective field theory, fictitious
ghost quarks are added to the Lagrangian. Ghost quarks have
same masses as valence quarks but are bosons, so the loop
with ghost quark goes
with an opposite sign and they cancel the valence contributions.
Then only the effects of the sea quarks are left.
Choosing different sea quark masses other than those of valence
quarks in the field theory with ghost quarks is therefore
equivalent to the partial quenching artifact in lattice QCD
simulations which use different masses for sea and valence quarks.
In the limit of $m^\mathrm{sea} = m^\mathrm{val}$, real QCD is recovered.

Partially quenched $\chi$PT (PQ$\chi$PT) and
quenched $\chi$PT are two effective field theories with
fictitious ghost fields.
The former is useful for chiral extrapolations in lattice calculations
to physical sea quark masses,
which are however missing in the latter theory.
In this section we will
focus on partially quenched chiral corrections only.
The major modification to the derivation in the
previous section is that there is a modified propagator for the
Goldstone mesons.

The PQ$\chi$PT pseudoscalar Goldstone meson sector is
described by the Lagrangian~\cite{Morel:1987xk,Sharpe:1992ft,Bernard:1992mk,
Bernard:1993sv,Sharpe:1997by,Golterman:1997st,Sharpe:2000bc,Sharpe:2001fh},
\begin{align}
  {\cal L} &=
    \frac{f^2}{8} \str \Big(
      \partial^\mu \S^\dagger \partial_\mu \S \Big)
      + \frac{\l}{4}  \, \str \Big( m_q \S^\dagger + m_q^\dagger \S \Big)
 \nl & \quad
           +\a_\Phi \partial^\mu \Phi_0 \partial_\mu \Phi_0
           - \mu_0^2 \Phi_0^2.
\label{eq:pqbosons}
\end{align}
Here $\str( )$ denotes the supertrace over flavor indexes, i.e.,
${\rm str}(A) = \sum_a \epsilon_a A_{aa}$, where $\epsilon_a = 1$
for $a=1, \cdots, 6,$ and $\epsilon_a = -1$ for $a=7,8,9$.
The field $\Sigma$ is defined by
\begin{equation}
  \S = \exp \Big( \frac{2 i \Phi}{f} \Big) = \xi^2,
\end{equation}
and the meson fields appear in the $U(6|3)$ matrix,
\begin{equation}
    \Phi =
    \begin{pmatrix}
      M & \chi^\dagger\\
      \chi & \tilde M\\
    \end{pmatrix}. \label{eq:mesonmatrix}
\end{equation}
$M$ and $\tilde M$ matrices contain bosonic mesons,
while  $\chi$ and $\chi^\dagger$ matrices contain fermionic mesons
(one ghost quark with one sea/valence quark).
The quark mass matrix is defined as~\cite{Mehen:2006vv}
\bea
m_q = {\rm diag}(m_u, m_d, m_s, m_j, m_l, m_r, m_u, m_d, m_s).
\eea

In the isospin limit, $m_u=m_d$ and $m_j=m_l$ for both the valence and sea sectors.
We set the strange sea quark mass to be same as
the valence quark mass , $m_r=m_s$. From the lowest order \PQCPT Lagrangian,
the meson with quark content $q \bar q'$ has the mass of
$m_{q \bar q'}^2 = \frac{\lambda}{f^2}(m_q +m_{q'})$.
The PQ$\chi$PT propagators of the off-diagonal mesons have the usual
Klein-Gordon form.
The flavor neutral propagator can be conveniently written
as~\cite{Mehen:2006vv}
\bea
{\cal G}^{PQ}_{a b} =
         \e_a \d_{ab} P_a +
         {\cal P}_{ab}\Big(P_a,P_b,P_X\Big),
\eea
where $ P_a = \frac{i}{q^2 - m^2_{aa} +i\e},
     P_b = \frac{i}{q^2 - m^2_{bb} +i\e},
     P_X = \frac{i}{q^2 - m^2_X +i\e}$,
     with $m_X^2 = \frac{1}{3}(m_{jj}^2+2m_{rr}^2)$
and
\begin{align}
     {\cal P}_{ab}\big(A,B,C\Big) &=
           -\frac{1}{3} \Bigg[
             \frac{\big( m^2_{aa} - m^2_{jj}\big)
                   \big( m^2_{aa} - m^2_{rr}\big)}
                  {\big( m^2_{aa} - m^2_{bb}\big)
                   \big( m^2_{aa} - m^2_X\big)}
                 A
    \nl & \quad
    +\frac{\big( m^2_{bb} - m^2_{jj}\big)
                   \big( m^2_{bb} - m^2_{rr}\big)}
                  {\big( m^2_{bb} - m^2_{aa}\big)
                   \big( m^2_{bb} - m^2_X\big)}
                 B
    \nl & \quad
            +\frac{\big( m^2_X - m^2_{jj}\big)
                   \big( m^2_X - m^2_{rr}\big)}
                  {\big( m^2_X - m^2_{aa}\big)
                   \big( m^2_X - m^2_{bb}\big)}
                 C \Bigg].
    \end{align}

In terms of the $5 \times 2$ field, the partially quenched Lagrangian
for doubly heavy baryon and heavy meson is~\cite{Mehen:2006vv}
\begin{align} \label{eq:LPQ}
\cL^{PQ} &=
\big( \fH^\dagger (\fH i \overset{\leftarrow}{D}_0 )\big)
-
g ( \fH^\dagger \fH \bm A \cdot \bm{\sigma})
+
\frac{\D_H}{4} ( \fH^\dagger \bm{\Sigma} \cdot \fH \bm{\sigma} )
\nl & \quad
+
\sigma ( \fH^\dagger \fH \cM )
+
\sigma' (\fH^\dagger \fH ) \str (\cM).
\end{align}
$\cM = \frac{1}{2} \big( \xi m_q \xi + \xi^\dagger m_q \xi^\dagger \big)$
is the mass operator.
 $\sigma$ and $\sigma' $
are the coupling constants in the mass operators.
The baryon mass splittings in PQ$\chi$PT are given by
\begin{align*}
\Delta_{ccqq'}
&=m_{\Xi_{\bar c \bar c \bar q'}}
    \!\!-m_{\Xi_{\bar c \bar c \bar q}}
 \!\!= m_{\Xi^{\ast}_{\bar c\bar c\bar q'}}
    \!\!-m_{\Xi^{\ast}_{\bar c \bar c\bar q}}
 \!\!= - \sigma (m_{q'}\!-m_q) ,
\nl
\Delta_{bcqq'}
&=m_{\Xi_{bcq'}}
    \!\!-m_{\Xi_{bcq}}
  \!\! = m_{\Xi^{\ast}_{\bar b\bar c \bar q'}}
    \!\!-m_{\Xi^{\ast}_{\bar b\bar c\bar q}}
  \!\! = - \sigma (m_{q'}-m_q) ,
\nl
\Delta^{\ast}_{ccqq'}
&= m_{\Xi^{\ast}_{\bar c\bar c\bar q'}}
    \!\!-m_{\Xi_{\bar c\bar c\bar q}}
  \!\!= \tfrac{3}{4} \Delta_H
    - \sigma (m_{q'}-m_q) ,
\nl
\Delta^{\ast}_{bcqq'}
&= m_{\Xi^{\ast}_{\bar b\bar c\bar q'}}
    \!\!-m_{\Xi_{\bar b\bar c\bar q}}
   \!\!= \tfrac{1}{2}\big(
\tfrac{3}{4} \Delta_H \big) - \sigma (m_{q'}-m_q) ,
\end{align*}
 with $\mu_Q = \frac{m_b m_c}{m_b+m_c} \approx m_c$.

Evaluating the PQ$\chi$PT loop diagrams, such as Fig.~\ref{oneloop},
we obtain PQ$\chi$PT corrections to the six form factors and list
the formulism in Appendix.
The experimental value of $SU(3)$ splitting of
the ground state $D$ mesons is
$m_{D_s} - m_D = - \sigma (m_s - m_u) \approx 100$
MeV.
It together with $m_\pi^2 = \frac{\lambda}{f^2}(m_u+m_d)$ and
$m_k^2 = \frac{\lambda}{f^2}(m_s+m_d)$ lead to
$\sigma = -\frac{100\, ({\rm MeV})}{m_k^2 - m_\pi ^2} \frac{\lambda}{f^2}$.
Other parameters are  same as in previous section
(cf.\ Table~\ref{tab:chicorr}).
Thus only three
parameters can be varied: the valence pion mass $m_{\pi}^{\mathrm{val}} = m_{uu}$,
the sea pion mass $m_{\pi}^{\mathrm{sea}} = m_{jj}$, and the renormalization
scale $\mu$.
\begin{figure}
\epsfig{file=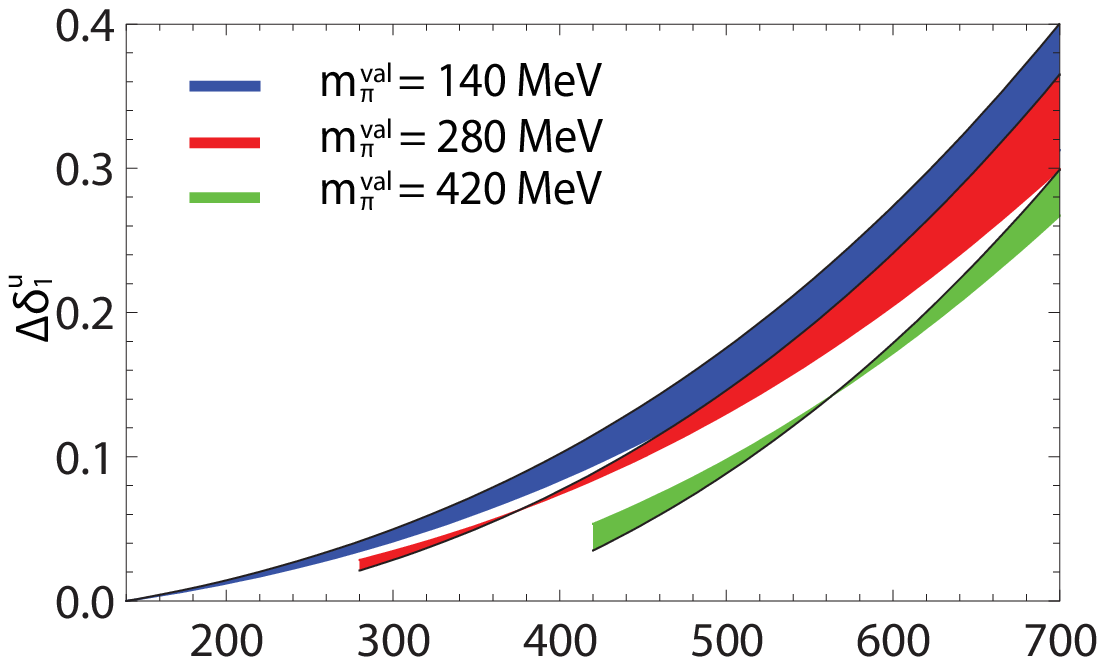, width=80mm}
\epsfig{file=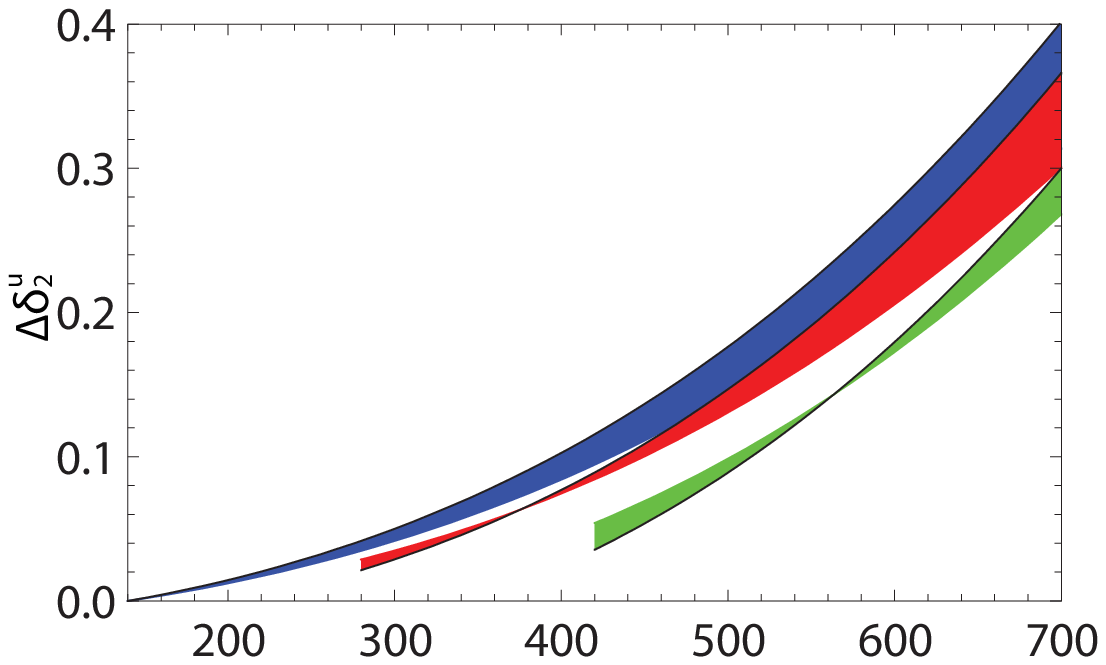, width=80mm}
\epsfig{file=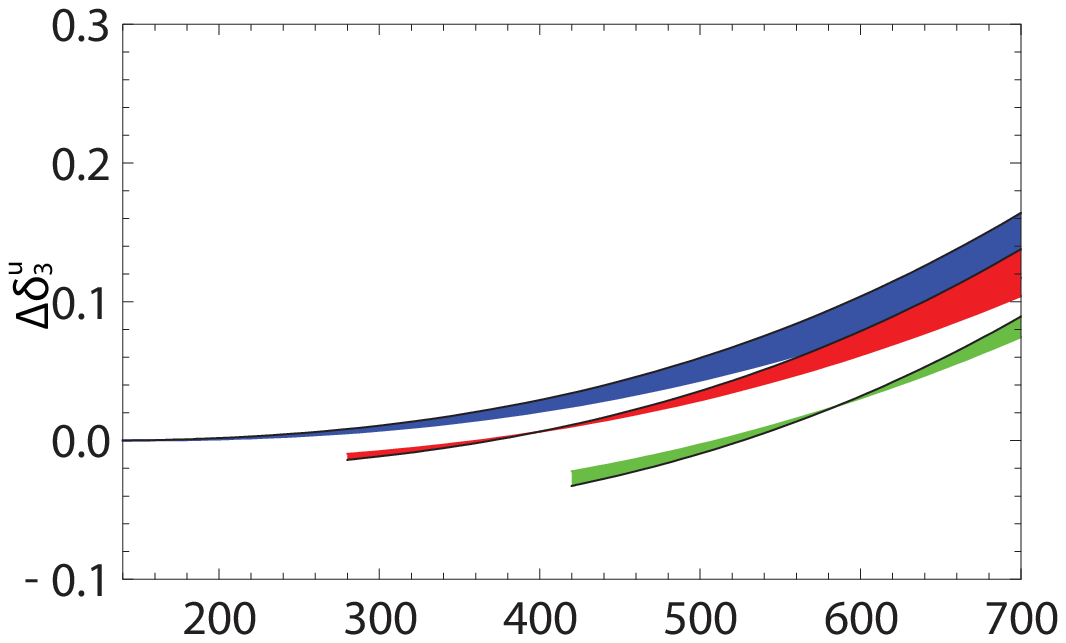, width=80mm}
\epsfig{file=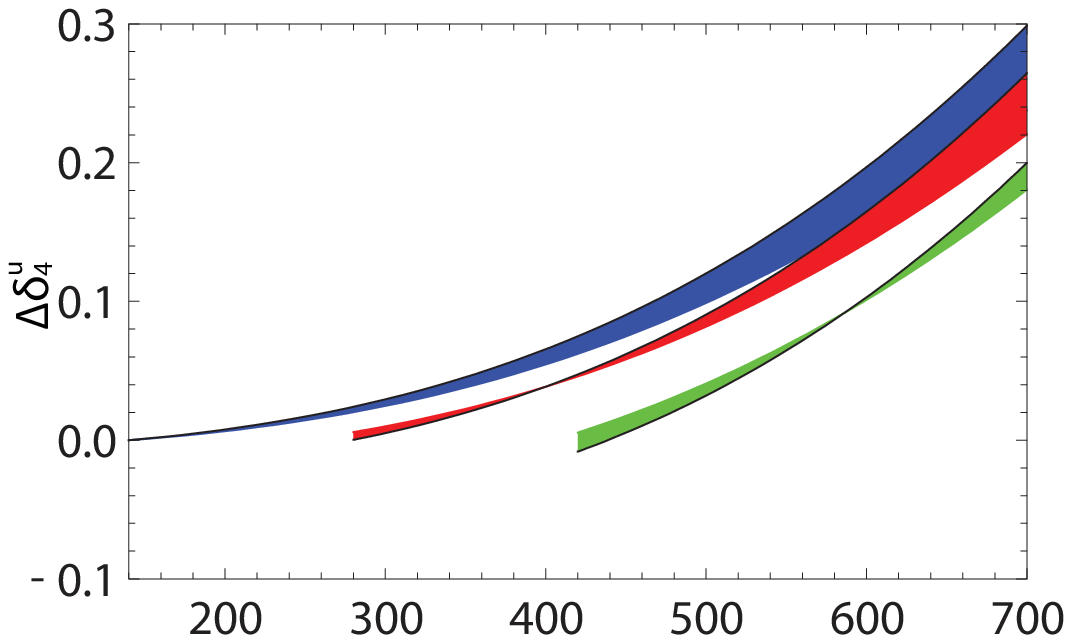, width=80mm}
\epsfig{file=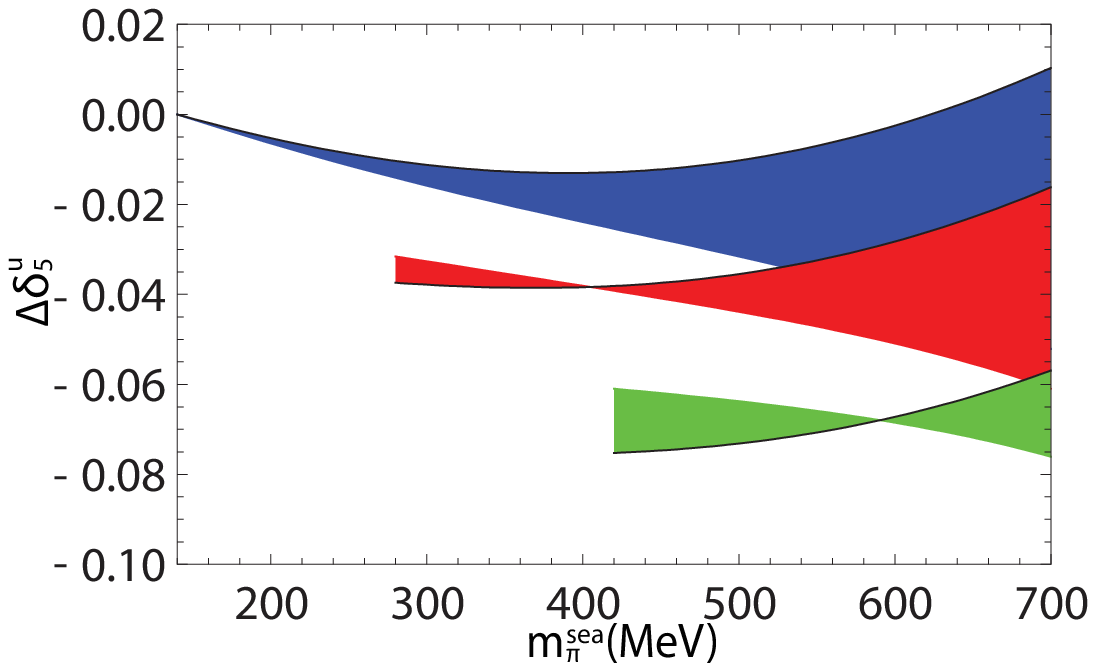, width=80mm}
\epsfig{file=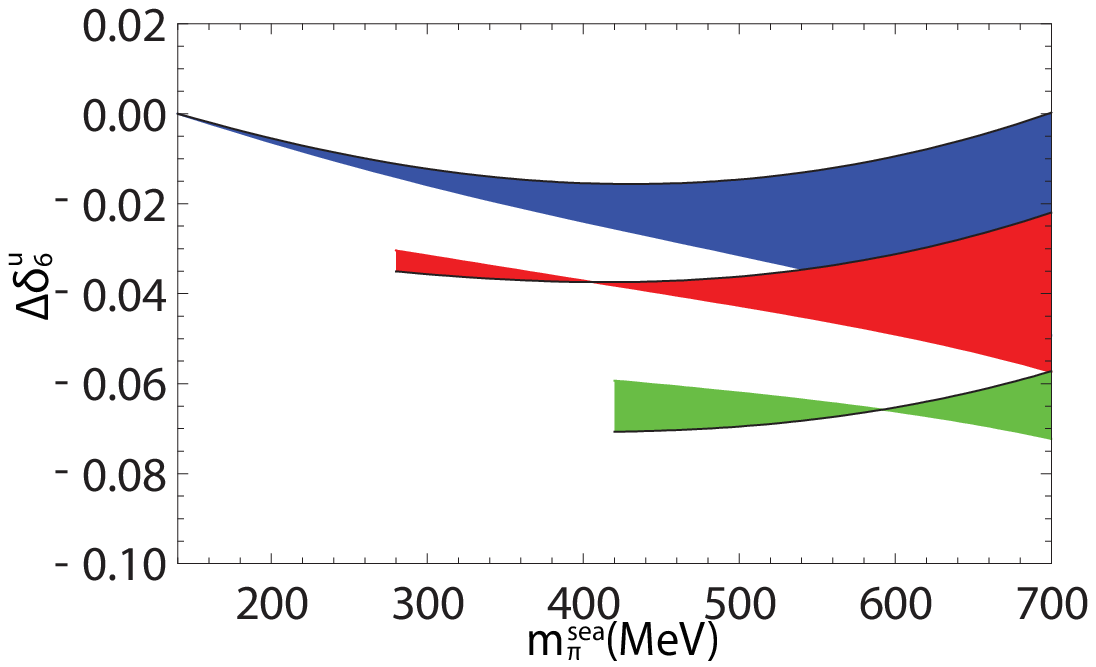, width=80mm}
 {\tighten
\caption[1]{$\Delta \d^u_{i, {\PQ}} $ as a function of
$m_{\pi}^{\mathrm{sea}} $ for different values of $m_{\pi }^{\mathrm{val}}$.
The width of the bands are the results of varying $\mu$
between $500$ MeV and $1500$ MeV. }
\label{u} }
\end{figure}
\begin{figure}
\centering
\epsfig{file=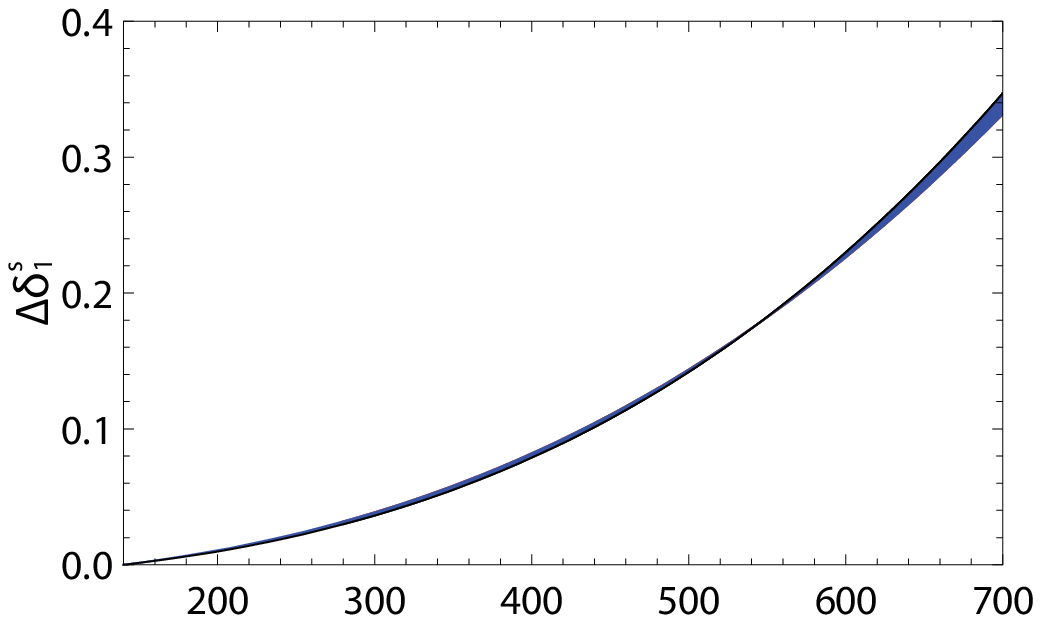, width=80mm}
\epsfig{file=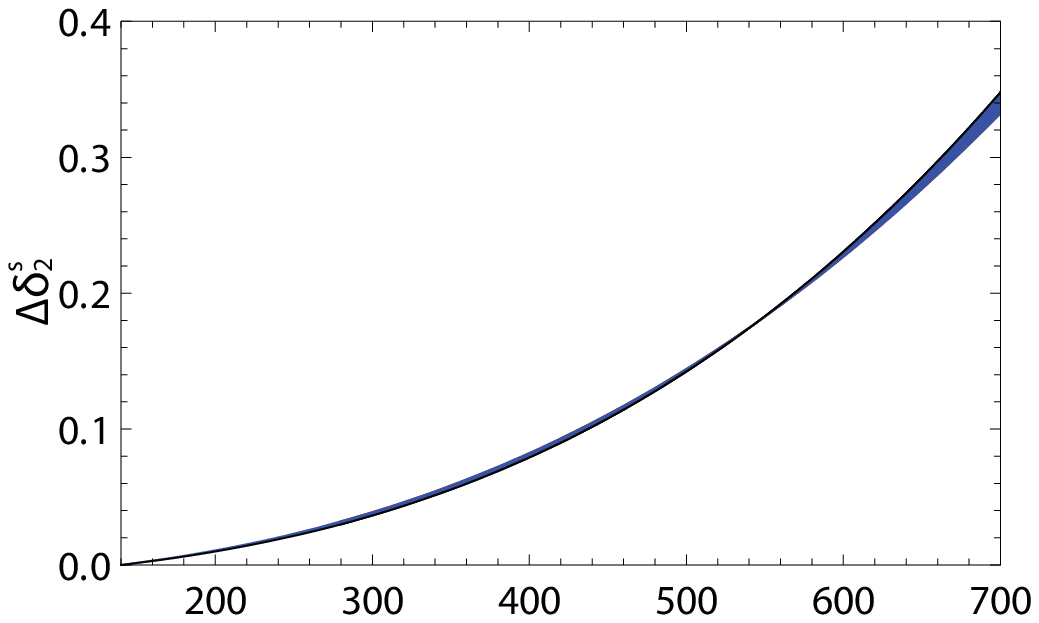, width=80mm}
\epsfig{file=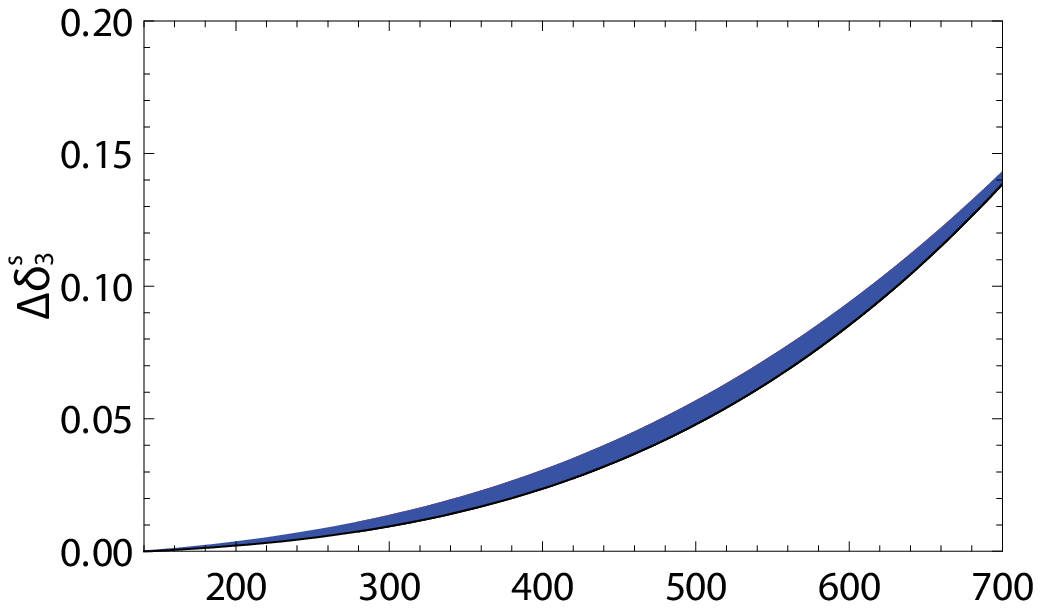, width=80mm}
\epsfig{file=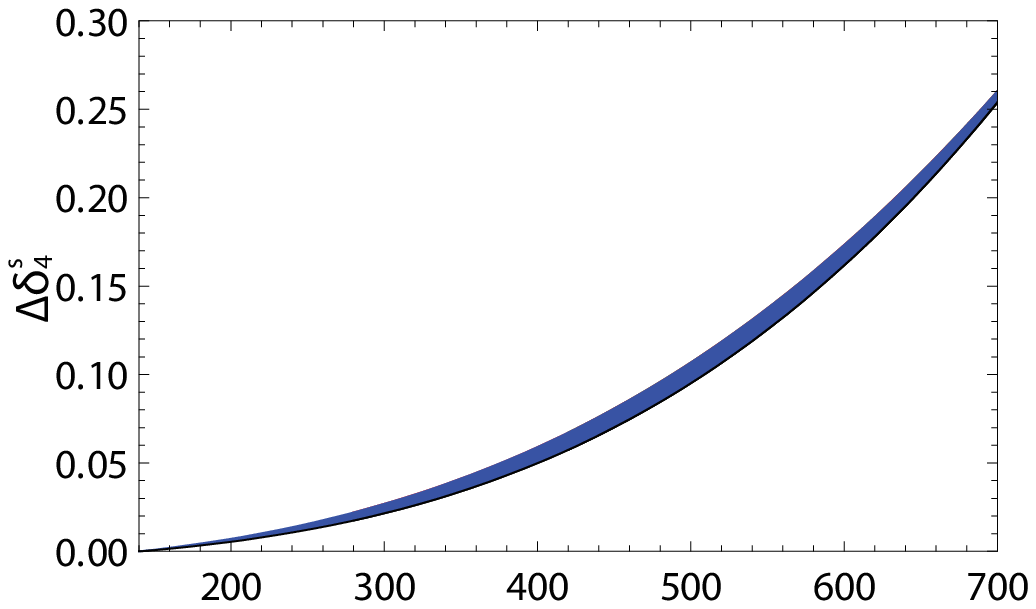, width=80mm}
\epsfig{file=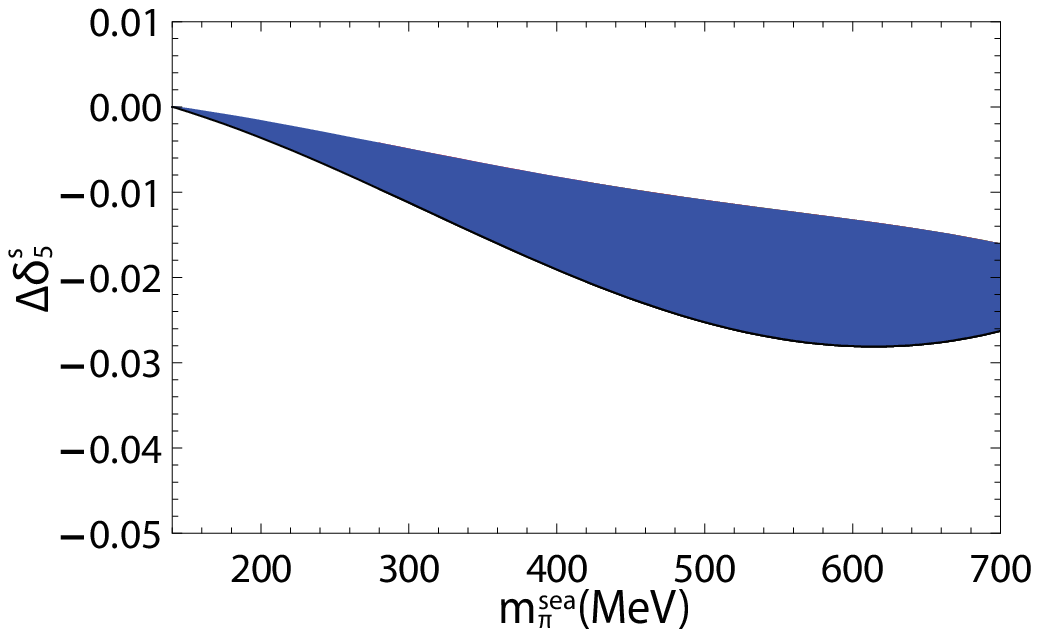, width=80mm}
\epsfig{file=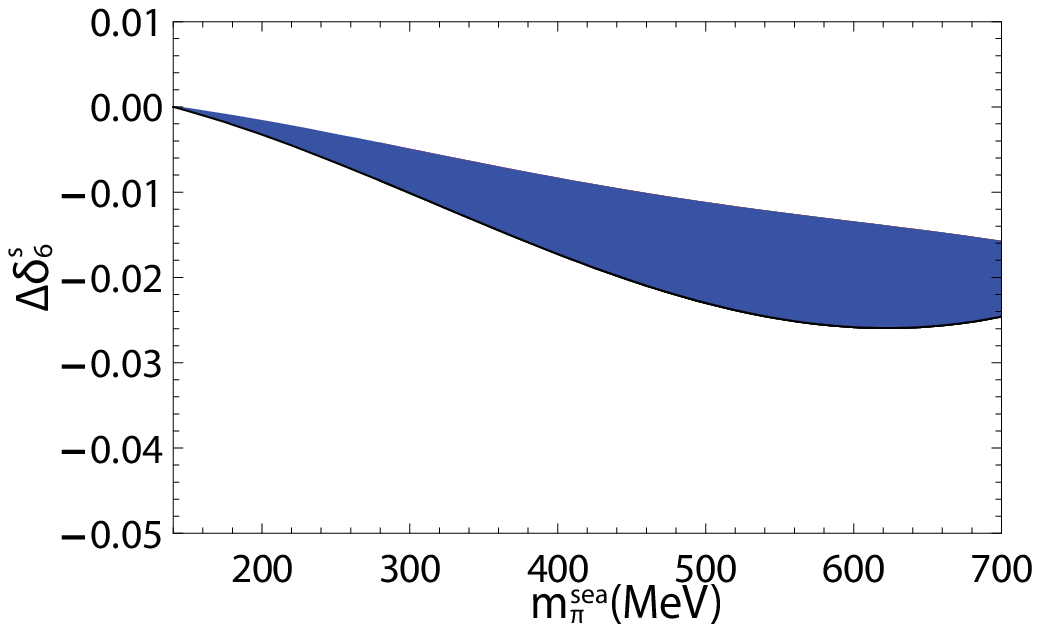, width=80mm}
 {\tighten
\caption[1]{$\Delta \d^s _{i, {\PQ}} $  as a function of
$m_{\pi}^{\mathrm{sea}} $ but independent of different values of $m_{\pi }^{\mathrm{val}}$.
The width of the bands are the results of varying $\mu$
between $500$ MeV and $1500$ MeV. }
\label{s} }
\end{figure}
In Figs.~\ref{u} and \ref{s} we plot the differences between \PQCPT corrections
and \CPT corrections, $\Delta \d^q_{i, \PQ} =\d^q_{i, \PQ} -  \d^q_{i}$,
in terms of $m_{\pi}^{\mathrm{sea}}$ with different values of $m_{\pi }^{\mathrm{val}}$
for both $q=u$ and $q=s$. For each value of
$m_{\pi }^{\mathrm{val}}=140$ MeV, $280$ MeV and $420$ MeV,
we let the $m_{\pi }^{\mathrm{sea}}$ range from $m_{\pi }^{\mathrm{val}}$ up to the mass
of eta-strange, $m_{\eta_s}= m_{ss}\approx 700$ MeV.
The bands correspond to varying $\mu$ from $500$ MeV to $1500$ MeV,
which is chosen to be the same for both $\chi$PT and PQ$\chi$PT.

From the plots, it is easy to see that the partially quenched
chiral corrections reproduce the chiral corrections when the
sea quark mass goes to the physical valence quark mass.
As demonstrated by the narrow bands in Figs.~\ref{u} and~\ref{s}, $\Delta \d^q_{i, \PQ}$ are
very insensitive to the choice of $\mu$.
The $\Delta \d^q_{1, \PQ} \sim \Delta \d^q_{4, \PQ}$
are affected most by sea quark mass and gain additional  partial quenching corrections
of range from $10\sim 40\%$.
Those corrections increase with increasing $m_{\pi}^{\mathrm{sea}}$ values.
The $\Delta \d^q_{5, \PQ} $ and $\Delta \d^q_{6, \PQ} $  of order $ 1\sim 8\%$
are very insensitive to partial quenching
effects and are insensitive to sea
quark mass.
It would be
interesting to test the chiral predictions in a lattice simulation.

\section{Summary} \label{sum}
In this paper we have used NRQCD to derive the coupling between heavy
diquarks and weak current with heavy quark-diquark symmetry. We
constructed the chiral Lagrangian for doubly heavy baryons coupled
to weak current and evaluated the tree level predictions for
doubly heavy baryon semileptonic weak decay form factors. We also
evaluated the chiral corrections to the form factors in both unquenched and partially
quenched theory and the formulism are given in Appendix. The partial quenching
formulae will be useful for chiral extrapolation of doubly heavy
baryon zero-recoil semileptonic decay form factors in lattice QCD
simulations. It will be interesting to test the calculations of
this paper with either experimental data or lattice simulations.

\begin{acknowledgments}

We thank B. Tiburzi at Maryland University for discussions and
especially thank T. Mehen at Duke University for various
discussions and kind guidance. This research is supported
in part by DOE grants DE-FG02-05ER41368 and DE-FG02-05ER41376.

 \end{acknowledgments}

\appendix*

\section{One-loop $\chi$PT and PQ$\chi$PT corrections to
zero-recoil semileptonic decay.}

The detailed one-loop $\chi$PT corrections $\d_{i}^q$ $(i=1,\cdots, 6)$
and PQ$\chi$PT corrections $\d_{i,\PQ}^q (i=1,\cdots, 6)$
to form factors of zero-recoil
semileptonic decay are provided in this section.
 $\overline{MS}$
scheme is used in this derivation
and unknown counterterms are not included.

In the following chiral correction
expressions, $f$ is the pion
decay constant, $m_i$
is the mass of the Goldstone boson in the one-loop diagram,
and ${C}^i_{ab}$ is a factor which arrives
from $SU(3)$ Clebsch-Gordan coefficients in the couplings.
 ${C}^{\pi^{\pm}}_{12} = {C}^{\pi^{\pm}}_{21} =1$
 for loops with charged pions,
${C}^{\pi^0}_{11} = {C}^{\pi^0}_{22} = \frac{1}{2}$
for loops with neutral pions,
${C}^{K}_{3i} = {C}^{K}_{i3} =1$ ($i =1$ or $2$)
for loops with kaons, and
$C^\eta_{11}=C^\eta_{22}= \frac{1}{6}, C^\eta_{33} =\frac{2}{3}$
for loops with $\eta$ mesons.

$\chi$PT corrections $\d_{i}^q$ $(i=1,\cdots, 6)$
are written in terms of function $I(\Delta_1, \Delta_2, m,\mu)$
which are defined later:
\begin{align}
\d_{1}^q &= - \tfrac{g^2}{(4 \pi f)^2}
 \sum_{i, q'} C^i_{qq'} \big[ \tfrac{1}{9}
I(\Delta_{bcqq'},\Delta_{ccqq'}, m_i, \mu )
\nl & \quad
+ \tfrac{8}{9} I(\Delta^{\ast}_{bcqq'},\Delta^{\ast}_{ccqq'}, m_i, \mu ) \big] \, ,
\nl
\d_{2}^q &= - \tfrac{g^2}{(4 \pi f)^2}
\sum_{i, q'} C^i_{qq'}
\big[ \tfrac{-1}{27} I(\Delta_{bcqq'},\Delta_{ccqq'}, m_i, \mu )
\nl & \quad
+ \tfrac{4}{27} I(\Delta_{bcqq'},\Delta^{\ast}_{ccqq'}, m_i, \mu )
\nl & \quad
+ \tfrac{4}{27} I(\Delta^{\ast}_{bcqq'},\Delta_{ccqq'}, m_i, \mu )
\nl & \quad
+ \tfrac{20}{27} I(\Delta^{\ast}_{bcqq'},\Delta^{\ast}_{ccqq'}, m_i, \mu ) \big] \, ,
\nl
\d_{3}^q &= -\tfrac{g^2}{(4 \pi f)^2}
 \sum_{i, q'} C^i_{qq'}
 \big[ \tfrac{8}{27}
I(\Delta_{bcqq'},- \Delta^{\ast}_{ccqq'}, m_i, \mu )
\nl & \quad
- \tfrac{5}{27} I(\Delta_{bcqq'},\Delta_{ccqq'}, m_i, \mu )
\nl & \quad
+ \tfrac{4}{27} I(\Delta^{\ast}_{bcqq'},-\Delta^{\ast}_{ccqq'}, m_i, \mu )
\nl & \quad
+ \tfrac{20}{27} I(\Delta^{\ast}_{bcqq'},\Delta_{ccqq'}, m_i, \mu ) \big] \, ,
\nl
\d_{4}^q &= -\tfrac{g^2}{(4 \pi f)^2}
 \sum_{i, q'} C^i_{qq'}
\big[ \tfrac{8}{27}
I(-\Delta^{\ast}_{bcqq'},\Delta_{ccqq'}, m_i, \mu )
\nl & \quad
- \tfrac{5}{27} I(-\Delta^{\ast}_{bcqq'},\Delta^{\ast}_{ccqq'}, m_i, \mu )
\nl & \quad
+ \tfrac{4}{27} I(\Delta_{bcqq'},\Delta_{ccqq'}, m_i, \mu )
\nl & \quad
+ \tfrac{20}{27} I(\Delta_{bcqq'},\Delta^{\ast}_{ccqq'}, m_i, \mu ) \big] \, ,
\end{align}
\begin{align}
\d_{5}^q &=-\tfrac{g^2}{(4 \pi f)^2}
  \sum_{i, q'} C^i_{qq'}
\big[\tfrac{4}{9}
I(-\Delta^{\ast}_{bcqq'},-\Delta^{\ast}_{ccqq'}, m_i, \mu )
\nl & \quad
+\tfrac{5}{9}  I(\Delta_{bcqq'},\Delta_{ccqq'}, m_i, \mu ) \big] \, ,
\nl
\d_{6}^q &= -\tfrac{g^2}{(4 \pi f)^2} \sum_{i, q'} C^i_{qq'}
\big[ \tfrac{8}{27}
I(-\Delta^{\ast}_{bcqq'},- \Delta^{\ast}_{ccqq'}, m_i, \mu )
\nl & \quad
+ \tfrac{4}{27} I(-\Delta^{\ast}_{bcqq'},\Delta_{ccqq'}, m_i, \mu )
\nl & \quad
+ \tfrac{4}{27}
I(\Delta_{bcqq'},- \Delta^{\ast}_{ccqq'}, m_i, \mu )
\nl & \quad
+ \tfrac{11}{27} I(\Delta_{bcqq'},\Delta_{ccqq'}, m_i, \mu ) \big] \, .
\end{align}
The functions $I$ are defined as
\begin{align}
I(0,0,m,\mu)\!&= \!0,
\nl
I(\Delta, \Delta, m, \mu)\!
&=\! -6 \Delta^2 \ln\Big(\tfrac{\mu^2}{m^2}\Big)
+ 4 \Big(m^2-3 \Delta^2 \Big)
\nl & \quad
+8 m\Delta F \!\Big(\tfrac{\Delta}{m}\Big)\!
+4 \Big(\Delta^2\!- m^2 \Big)F'\!\Big(\tfrac{\Delta}{m}\Big), \nl
I(\Delta _1, \Delta _2, m, \mu) \!
&=\! -2\Big(\Delta_1^2 +\Delta_1 \Delta_2+\Delta_2^2 \Big)
\ln\Big(\tfrac{\mu^2}{m^2}\Big)
\nl & \quad
+4 \Big(m^2-\Delta_1^2-\Delta_1 \Delta_2-\Delta_2^2 \Big)
\nl & \quad
+ \tfrac{4}{\Delta_2-\Delta_1} \Big[ (\Delta_2^2- m^2 )F\!
\Big(\tfrac{\Delta_2}{m}\Big) m
\nl & \quad \quad
-
(\Delta_1^2- m^2 )F\!\Big(\tfrac{\Delta_1}{m}\Big) m \Big],
\end{align}
where
\bea
F(x) \!= \!\left\{
\begin{array}{cc}
\!\!-\sqrt{1-x^2} \Big( \frac{\pi}{2}
\!-\! \rm{tan} ^{-1}\big(\frac{x}{\sqrt{1-x^2}}\big) \Big), &\! |x|\!<\!1 \\
\!\! \sqrt{x^2-1}\ln\Big(x \!+\! \sqrt{x^2-1}\Big),  & \! |x|\!\ge\!1
\end{array},\right.
\eea
and $F'(x)$ is the first derivative of $x$.

PQ$\chi$PT corrections $\d_{i,\PQ}^q$ $(i=1,\cdots, 6)$
are written in terms of function
 $I(\Delta_1, \Delta_2, m,\mu)$
and $K(\Delta_1, \Delta_2, m, m,\mu)$:

\begin{align}
\d_{1,\PQ}^q &= -\tfrac{g^2}{(4 \pi f)^2}
 \sum_{q'=j,l,r}
\big[ \tfrac{1}{9}
I(\Delta_{bcqq'},\Delta_{ccqq'}, m_{qq'}, \mu )
\nl & \quad
+ \tfrac{8}{9} I(\Delta^{\ast}_{bcqq'},\Delta^{\ast}_{ccqq'}, m_{qq'}, \mu )\big]
\nl & \quad
-\tfrac{g^2}{(4 \pi f)^2}\big[ \tfrac{1}{9}
K(\Delta_{bcqq},\Delta_{ccqq}, m_{qq}, m_{qq}, \mu )
\nl & \quad
+ \tfrac{8}{9} K(\Delta^{\ast}_{bcqq},\Delta^{\ast}_{ccqq}, m_{qq}, m_{qq}, \mu )\big],
\nl
\d_{2,\PQ}^q &= -\tfrac{g^2}{(4 \pi f)^2} \sum_{q'=j,l,r}
\big[ \tfrac{-1}{27} I(\Delta_{bcqq'},\Delta_{ccqq'}, m_{qq'}, \mu )
 \nl & \quad
 + \tfrac{4}{27} I(\Delta_{bcqq'},\Delta^{\ast}_{ccqq'}, m_{qq'}, \mu )
\nl & \quad
+ \tfrac{4}{27} I(\Delta^{\ast}_{bcqq'},\Delta_{ccqq'}, m_{qq'}, \mu )
\nl & \quad
+ \tfrac{20}{27} I(\Delta^{\ast}_{bcqq'},\Delta^{\ast}_{ccqq'}, m_{qq'}, \mu ) \big]
\nl & \quad
-\tfrac{g^2}{(4 \pi f)^2} \big[ \tfrac{-1}{27}
K(\Delta_{bcqq},\Delta_{ccqq}, m_{qq},m_{qq},  \mu )
\nl & \quad
+ \tfrac{4}{27} K(\Delta_{bcqq},\Delta^{\ast}_{ccqq}, m_{qq}, m_{qq}, \mu )
\nl & \quad
+ \tfrac{4}{27} K(\Delta^{\ast}_{bcqq},\Delta_{ccqq}, m_{qq},m_{qq},  \mu )
\nl & \quad
+ \tfrac{20}{27} K(\Delta^{\ast}_{bcqq},\Delta^{\ast}_{ccqq}, m_{qq}, m_{qq}, \mu ) \big] ,
\end{align}
\begin{align}
\d_{3,\PQ}^q &= - \tfrac{g^2}{(4 \pi f)^2} \sum_{q'=j,l,r}
\big[ \tfrac{8}{27} I(\Delta_{bcqq'},- \Delta^{\ast}_{ccqq'}, m_{qq'}, \mu )
\nl & \quad
- \tfrac{5}{27} I(\Delta_{bcqq'},\Delta_{ccqq'}, m_{qq'}, \mu )
\nl & \quad
+ \tfrac{4}{27} I(\Delta^{\ast}_{bcqq'},-\Delta^{\ast}_{ccqq'}, m_{qq'}, \mu )
\nl & \quad
+ \tfrac{20}{27} I(\Delta^{\ast}_{bcqq'},\Delta_{ccqq'}, m_{qq'}, \mu ) \big]
\nl & \quad
-  \tfrac{g^2}{(4 \pi f)^2} \big[ \tfrac{8}{27}
K(\Delta_{bcqq},- \Delta^{\ast}_{ccqq}, m_{qq},m_{qq},  \mu )
\nl & \quad
- \tfrac{5}{27}
K(\Delta_{bcqq},\Delta_{ccqq}, m_{qq},m_{qq},  \mu )
\nl & \quad
+ \tfrac{4}{27}
K(\Delta^{\ast}_{bcqq},-\Delta^{\ast}_{ccqq}, m_{qq},m_{qq},  \mu )
\nl & \quad
+ \tfrac{20}{27} K(\Delta^{\ast}_{bcqq},\Delta_{ccqq}, m_{qq}, m_{qq}, \mu ) \big],
%
\nl
\d_{4,\PQ}^q &= -\tfrac{g^2}{(4 \pi f)^2} \sum_{q' = j,l,r}
\big[ \tfrac{8}{27} I(-\Delta^{\ast}_{bcqq'},\Delta_{ccqq'}, m_{qq'}, \mu )
\nl & \quad
- \tfrac{5}{27} I(-\Delta^{\ast}_{bcqq'},\Delta^{\ast}_{ccqq'}, m_{qq'}, \mu )
\nl & \quad
+ \tfrac{4}{27} I(\Delta_{bcqq'},\Delta_{ccqq'}, m_{qq'}, \mu )
\nl & \quad
+ \tfrac{20}{27} I(\Delta_{bcqq'},\Delta^{\ast}_{ccqq'}, m_{qq'}, \mu ) \big]
\nl & \quad
-\tfrac{g^2}{(4 \pi f)^2}\big[ \tfrac{8}{27}
K(-\Delta^{\ast}_{bcqq},\Delta_{ccqq}, m_{qq},m_{qq},  \mu )
\nl & \quad
- \tfrac{5}{27}
K(-\Delta^{\ast}_{bcqq},\Delta^{\ast}_{ccqq}, m_{qq}, m_{qq}, \mu )
\nl & \quad
+ \tfrac{4}{27}
K(\Delta_{bcqq},\Delta_{ccqq}, m_{qq}, m_{qq}, \mu )
\nl & \quad
+ \tfrac{20}{27} K(\Delta_{bcqq},\Delta^{\ast}_{ccqq}, m_{qq},m_{qq},  \mu ) \big] ,
\nn
\end{align}
%
\begin{align}
\d_{5,\PQ}^q &= - \tfrac{g^2}{(4 \pi f)^2} \sum_{q'= j,l,r}
\big[\tfrac{4}{9} I(-\Delta^{\ast}_{bcqq'},-\Delta^{\ast}_{ccqq'}, m_{qq'}, \mu )
\nl & \quad
+\tfrac{5}{9}  I(\Delta_{bcqq'},\Delta_{ccqq'}, m_{qq'}, \mu ) \big]
\nl & \quad
-  \tfrac{g^2}{(4 \pi f)^2} \big[\tfrac{4}{9}
K(-\Delta^{\ast}_{bcqq},-\Delta^{\ast}_{ccqq}, m_{qq}, m_{qq}, \mu )
\nl & \quad
+\tfrac{5}{9}  k(\Delta_{bcqq},\Delta_{ccqq}, m_{qq}, m_{qq}, \mu ) \big],
\end{align}
\begin{align}
\d_{6,\PQ}^q &= - \tfrac{g^2}{(4 \pi f)^2} \sum_{q'=j,l,r}
\big[ \tfrac{8}{27} I(-\Delta^{\ast}_{bcqq'},- \Delta^{\ast}_{ccqq'}, m_{qq'}, \mu )
\nl & \quad
+ \tfrac{4}{27} I(-\Delta^{\ast}_{bcqq'},\Delta_{ccqq'}, m_{qq'}, \mu ) \,
\nl & \quad
+ \tfrac{4}{27}
I(\Delta_{bcqq'},- \Delta^{\ast}_{ccqq'}, m_{qq'}, \mu )
\nl & \quad
+ \tfrac{11}{27} I(\Delta_{bcqq'},\Delta_{ccqq'}, m_{qq'}, \mu ) \big]
\nl & \quad
- \tfrac{g^2}{(4 \pi f)^2}\big[\tfrac{8}{27}
K(-\Delta^{\ast}_{bcqq},- \Delta^{\ast}_{ccqq}, m_{qq},m_{qq},  \mu )
\nl & \quad
+ \tfrac{4}{27} K(-\Delta^{\ast}_{bcqq},\Delta_{ccqq}, m_{qq}, m_{qq}, \mu )
\nl & \quad
+ \tfrac{4}{27}
K(\Delta_{bcqq},- \Delta^{\ast}_{ccqq}, m_{qq},m_{qq},  \mu )
\nl & \quad
+ \tfrac{11}{27} K(\Delta_{bcqq},\Delta_{ccqq}, m_{qq},m_{qq},  \mu ) \big].
\end{align}
The function $K(\Delta_1, \Delta_2, m, m,\mu)$ which
arises from the hairpins is given by
\begin{align}
&K(\Delta_1, \Delta_2, m_a,m_b, \mu ) \nl
&\quad= \mathcal P_{ab}\big[I (\Delta_1, \Delta_2, m_a,\mu),
I (\Delta_1, \Delta_2, m_b,\mu) ,
\nl & \quad
\qquad  \quad I (\Delta_1, \Delta_2, m_X,\mu) \Big].
\end{align}

\end{document}